%% file: main.tex
\documentclass[conference]{IEEEtran}
\usepackage[utf8]{inputenc}
\usepackage{subfigure,times,graphics,epsfig,amsmath,xspace,endnotes,pifont,multirow,rotating,listings,amssymb,algorithmic,color,caption,nicefrac,adjustbox,todonotes,tabularx,mathtools, algorithmic,algorithm}
\usepackage{graphicx}
\usepackage{amsthm}
\newtheorem{thm}{Theorem}

\newtheorem{defn}[thm]{Definition}
\usepackage{dsfont}
\usepackage{tabularx}
\newcolumntype{L}[1]{>{\raggedright\arraybackslash}p{#1}} 
\newcolumntype{C}[1]{>{\centering\arraybackslash}p{#1}} 
\newcolumntype{R}[1]{>{\raggedleft\arraybackslash}p{#1}} %

\usepackage{amsmath}
\usepackage{amssymb}
\usepackage{accents}

\begin{document}
%
\title{Joint Satellite Gateway Deployment \& Controller Placement in Software-Defined 5G-Satellite Integrated Networks}

\author{\IEEEauthorblockN{Nariman Torkzaban}
\IEEEauthorblockA{Department of Electrical\\
and Computer Engineering \&}
\IEEEauthorblockA{Institute for Systems Research\\
University of Maryland\\
College Park, Maryland, USA\\
narimant@umd.edu}
\and
\IEEEauthorblockN{John S. Baras}
\IEEEauthorblockA{Department of Electrical \\
and Computer Engineering \&}
\IEEEauthorblockA{Institute for Systems Research\\
University of Maryland\\
College Park, Maryland, USA\\
baras@isr.umd.edu}}


 \maketitle

\begin{abstract}
Several challenging optimization problems arise while considering the deployment of the space-air-ground integrated networks (SAGINs), among which the optimal satellite gateway deployment problem is of significant importance.  Moreover, with the increasing interest in the software-defined integration of 5G networks and satellites, the existence of an effective scheme for optimal placement of SDN controllers, is essential. In this paper, we discuss the interrelation between the two problems above and propose suitable methods to solve them under various network design criteria. We first provide a MILP model for solving the joint problem, and then motivate the decomposition of the model into two disjoint MILPs. We then show that the resulting problems can be modeled as the optimization of submodular set functions and can be solved efficiently with provable optimality gaps.



\end{abstract}


%

\input{introduction1.tex}
\input{related_work.tex}
\input{problem_description.tex}
\input{formulations.tex}

\input{evaluation.tex}
\input{conclusion.tex}




%

\end{document}

%% file: introduction1.tex
\section{Introduction}

5G mobile communication systems are required to achieve key performance indicators (KPIs) in terms of low latency, massive device connectivity, consistent quality of service (QoS) and high security. For instance, user bit rates up to $10$ Gbps and round-trip times (RTTs) as small as $1 – 10$ ms are demanded in specific application scenarios in 5G. Moreover, due to the significantly increasing traffic volume, number of registered users, and new provisioned use-cases such as cloud computing, Internet of Things (IoT), massive-data applications, etc., it has become evident over the past few years that towards achieving the 5G key promises, it is essential to take advantage of the full capacity of all communications types \& segments (e.g. terrestrial, aerial, and space) as well as supporting technologies (e.g. SDN, NFV, etc.) simultaneously, otherwise the traditional stand-alone terrestrial networks will fail to achieve the key projected promises.

Based on the above discussion, towards the inevitable convergence of the above paradigms and technologies, an optimized integration and configuration policy, that is tailored to specific use-cases and corresponding QoS requirements, becomes of paramount importance. Moreover, given that by the end of $2020$, only less than $50\%$ of the world are projected to have access to stable communications, the importance of adopting and integrating satellite communications, has been recognised and supported by standardization bodies such as 3GPP, ITU \cite{itu}, and ETSI \cite{etsi}.

\begin{figure*}[t]
\begin{center}
\begin{minipage}[h]{0.925\textwidth}
\includegraphics[width=1\linewidth]{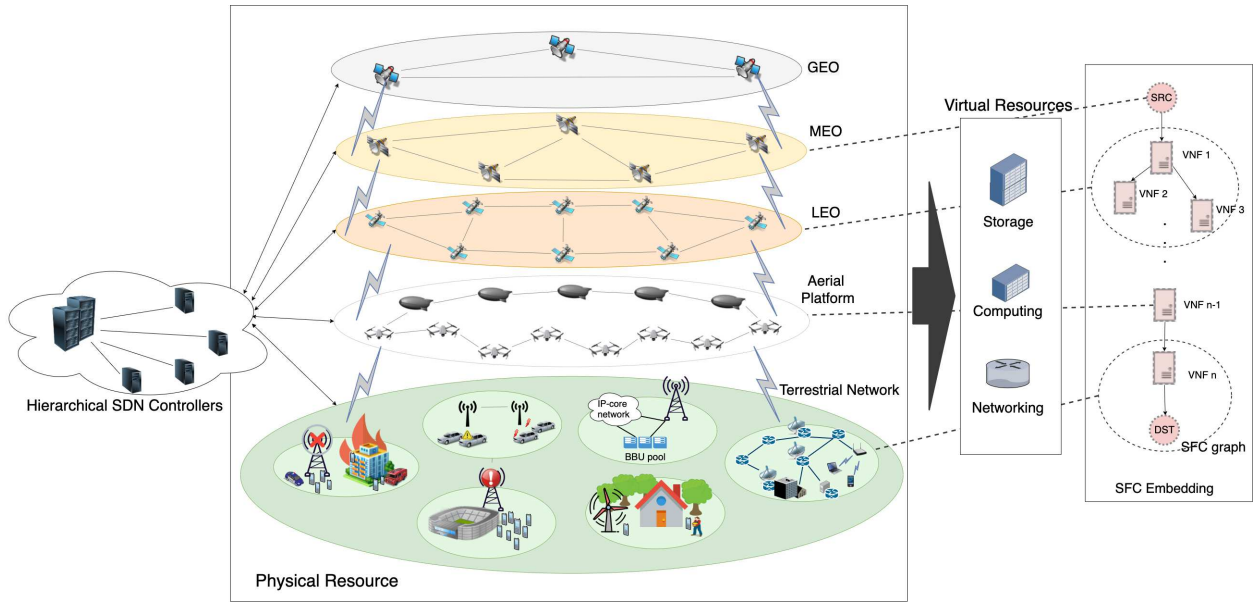}
\caption{ Space-Air-Ground Integrated Network (SAGIN)}
\label{fig:scene}
\end{minipage}
\hspace{1em}
\end{center}
\end{figure*}

The space-air-ground integrated network (SAGIN) \cite{katosurvey}, depicted in \ref{fig:scene} offers potential benefits which are not possible otherwise, including global coverage, low latency and high reliability. In particular, satellites can replace, extend, or complement the terrestrial networks, in rural, and hard-to-reach areas, or where the existence of communications infrastructure is costly or even infeasible in use-cases such as mountains and marine communications. Furthermore, satellites can offload the terrestrial networks by accommodating the delay-insensitive applications, allowing the terrestrial segment to survive when there is a surge in the traffic load. Finally, due to their global coverage, satellites can provide a reliable and seamless back-haul for aerial segment, and also the monitoring and control applications for IoT, vehicular networks, etc.

On the other hand, despite providing the advantages of the three different segments, new challenges are also introduced in the integrated network due to the limitations of each of the layers, including but not limited to, complicated end-to-end resource provisioning due to the additional resource constraints, high control complexity due to the different dynamics of each segment, non-unified interfaces between the layers, etc. which significantly impact the decisions regarding traffic routing, spectrum allocation, mobility management, QoS and traffic management, etc. These challenges, together with the diversity of 5G use-cases with large-scale applications, highlights the importance of a unified management and control structure, and a dynamic resource allocation policy which are both scalable and flexible enough to handle the increasing complexity. The key to address these issues is in the co-existence of two different but complementary supporting technologies namely,  Software-Defined Networking (SDN), and Network Function Virtualization (NFV).

SDN, allows for separating the control logic of the network from its forwarding logic and realizes a centralized management policy. This not only allows for a simple realization of the forwarding layer, but also paves the way for dynamic configuration of control and management policies. With NFV, the network functions are decoupled from the proprietary hardware and realized though software. This both reduces the hardware cost and also by implementing the network functions on virtual machines or containers, provides an easier and more flexible provisioning of scalable solutions driving higher profitability for the
network providers.



Towards realizing the SDN/NFV-enabled SAGIN Several important optimization problems arise immediately; i) In the architectures concerned with GSO satellites, due to high delivered throughput per satellite, large number of gateways are required; sometimes exceeding a couple of dozens. ii) Moreover, once the gateway deployment policy is decided, it is of paramount importance to develop a smart and adaptive mechanism to handle the user hand-overs between the gateways or LEO satellites, traffic routing, load balancing, etc. Due to their abstract view of the network, SDN controllers are the best fit for this purpose. Thus, it becomes essential to formulate an optimization problem for deciding the minimum number of gateways and SDN controllers and their optimal location within the SAGIN; iii) Furthermore, by the virtue of NFV, the complex network services corresponding to different applications in SAGIN provision, are modeled as service function chains (SFC)s (i.e. sets of ordered virtual functions with logical dependencies). The embedding of these SFCs on the SAGIN infrastructure, and establishing the traffic routes between them is another challenging problem.  

While approaching the above-mentioned problems, depending on the design goals, multiple objectives such as cost minimization, load balancing, mission offloading, reliability maximization, etc. can be sought under various network design constraints, such as guarantee requirements on end-to-end latency, network security or availability, fault tolerance, etc \cite{sfc}. Within the context of security-intensive applications, such as first-responders, autonomous vehicles, and tactical networks, reliability and trust plays an important role. So, a trust-aware mechanism needs to be taken in all decisions and interactions such as routing \cite{path-based}, service deployment \cite{trust-aware}, and computing. Moreover, given the logically centralized and physically distributed nature of the control plane it is important to ensure that the freshness of received network data is maximized \cite{javani} at the SDN controllers to guarantee a valid global view of the network. 

Within the framework of SAGIN, the control plane is mainly in charge of making the routing decisions, conducting the traffic hand-offs between the satellite gateways and satellite switches, ensuring the service QoS,  and delivering the necessary instructions to the SDN-enabled switches, leaving the only the simple forwarding task to the data plane switches. Given the central role of the SDN controllers in SDN-enabled SAGIN, it becomes important to maintain a reliable communication path between the SDN controllers with themselves and the SDN-enabled switches. 

The major contributions of this paper are as follows: 

\begin{itemize}
    \item We model the joint satellite gateway and SDN controller placement (JGCP) problem as a MILP. We consider two variants of this problem with two different sets of objectives. $(i)$ Jointly maximizing the reliability of network-to-gateway and network-to-controller assignments. $(ii)$ Minimizing the synchronization cost of the SDN controllers, jointly with minimizing the latency of the network-to-gateway and network-to-controller assignments. 
    
    \item Inspired by \cite{tass}, we provide a realistic scheme for modelling the synchronization overhead between the SDN controllers given the location and the network segment (terrestrial or space) in which each of the SDN controllers resides. 
    
    \item We decompose the MILP model into two disjoint MILPs and then show that the resulting models lie in the framework of submodular optimization. Then we apply two approximation methods for solving these two models that run efficiently in time and provide provable theoretical optimality gaps. 
    
    \item We conduct extensive experimental tests to evaluate the performance of the provided methods and algorithms. We use publicly available real-world scenarios and various simulation settings for the performance evaluation tasks. 
\end{itemize}

The remainder of the paper is organized as follows. Section~\ref{sec:desc} describes the network model and the problem description. In Section~\ref{sec:problem} we introduce our MILP formulations and provide suitable approximations to solve them. Section \ref{sec:evaluation} presents our evaluation results, whereas Section \ref{sec:relatedwork} provides an overview of related work. Finally, in Section \ref{sec:conclusions}, we highlight our conclusions and discuss directions for future work.

%% file: related_work.tex
\section{Related Work}
\label{sec:relatedwork}
 Although the gateway placement problem over a general network has been studied extensively, the satellite gateway deployment and the SDN controller placement problems in SDSNs are relatively new problems. In \cite{}, the authors have very well explained why the traditional gateway placement methods are not suitable to satisfactorily address these problems. Over the past few years there have been interesting works concerning these two problems in different settings and from several perspectives; The gateway placement with GEO satellites has received a lot of attention while the attention towards LEO constellations has increased a lot only in the past few years. Some works have primarily considered the optimization of the ground segment by optimizing the deployment of ground stations, while some other works were only concerned about the deployment of gateways and controllers on the aerial or space layer. 
 
 Several papers have studied gateway diversity towards the optimization of High-Throughput Satellite (HTS) architectures. These works typically aim at optimizing the smart gateway configuration \cite{ehf}, gateway assignment\cite{yunan}, paving the way towards optical feeder links and dealing with atmospheric turbulence \cite{ofl}. 
 
 Several works have been concerned with the gateway deployment and SDN controller placement in LEO constellations. In \cite{leog}, the authors have proposed a genetic algorithm for deployment of multiple gateways in a large constellation, taking into account the gateway-satellite connectivity, hop count, and the potential location of gateways, to achieve satisfactory performance in terms of load balance, latency, and traffic peak, towards an optimized layout. A joint gateway assignment and routing method is proposed in \cite{avoid} with the goal of congestion avoidance in mega-constellations. With respect to SDN controller placement, the research works differ mostly in the location of placing the controllers each of which providing some benefits and some shortcomings. Some papers decide to place the controllers on the ground segment, some place them on the LEO SDN-enabled satellite switches \cite{wolf}\cite{wolf2}, and some on the GEO layer \cite{geoc}, while some other works propose hierarchical controller architectures comprising ground stations, LEO, MEO, and GEO-layer controllers \cite{hier}, \cite{hierm}, \cite{hier3}.  
 In \cite{hierm}, \cite{wolf}, and, \cite{wolf2}, the authors consider the controller placement problem in both static and dynamic modes, where in the former the controller placement and satellite-to-controller assignments remain unchanged, while in the latter the number of the controllers, their locations, and therefore the assignments vary with respect to change in demand and traffic pattern over time. Further in \cite{wolf}, and \cite{wolf2}, the flow setup time is adopted as a metric which makes the problem statement realistic as optimizing the flow setup time is a major concern in SDN-enabled networks. 

Within the context of SAGIN, various objectives can be sought when solving the satellite gateway deployment problem that may be inspired by different requirements such as cost or latency minimization, reliability-awareness, route optimization, load balancing, etc.. The problem is usually formulated in the literature as a combinatorial optimization problem and is solved exactly or approximately. The authors in \cite{kato17}, aim at minimizing the average user-to-gateway latency over the network, given had reliability constraints, however, they only take into account the shortest paths from user to gateway and therefore no routing scheme is provided. They solve the problem using a Particle Swarm Optimization (PSO) algorithm and compare their approach to the brute-force greed search method in terms of time complexity. In \cite{onlyrel}, only maximization of reliability is considered, where the authors use a simple clustering algorithm and compare their method with the result of optimal enumeration to justify the lower time complexity at the expense of sub-optimal results. 

The joint gateway deployment and controller placement in SAGIN has also received an increasing attention over the past few years. The authors in \cite{scgp}, formulate a joint deployment of satellite gateways and SDN controllers to maximize the average reliability with hard constraints on user-to-satellite delay. They propose an iterative approach which relying on simulated annealing and clustering, where in each iteration first the current gateway placement policy and then the controller placement is updated towards the convergence. In \cite{part}, the exact same problem under similar settings and with similar objectives has been taken with the only difference that  the simulated annealing approach from \cite{scgp} is augmented with a portioning phase (separately w.r.t gateways \& controllers) to render several sub-problems of smaller size. Finally, in \cite{meta}, a number of meta-heuristic approaches namely, simulated annealing, double simulated annealing, and genetic algorithm for the same problem has been considered and their performance is compared.

Despite incorporating many interesting ideas, most of the mentioned works assume that the optimal number of equipment to deploy is known a-priori, do not consider the routing overhead and gateway load, and do not employ any realistic scheme for the control plane. Although these simplifications may render more tractable problems allowing for more interesting and smart solution, but are prone to become unusable in some real-world scenarios. 

In \cite{capac}, this issue is mitigated to some extent as the authors take into account the capacity constraints of satellite links and formulate a problem towards maximizing the average reliability which they solve using a greedy algorithm. In \cite{nar}, a joint gateway placement and traffic routing is formulated as an instance of the capacitated facility location-routing problem which does not make any assumption on the optimal number of gateways, and also provides a routing solution that does not contradict the capacity constraints and is not limited to the shortest paths using multi-commodity flow allocation.






%% file: problem_description.tex
\section{System Model} 
\label{sec:desc}

\subsection{Network Architecture}

We consider an SDN-enabled hybrid 5G-satellite network as in Fig. \ref{fig:scene} that consists of two logical segments; the \textit{data plane}, and the \textit{control plane}. The backhaul SDN-enabled switches reside within the data plane, where the users deliver their generated traffic to 5GC through the  Radio Access Network (RAN); i.e. gNBs, small cells, etc. The high-throughput GEO satellite is also an SDN-enabled switch. The control plane realizes a logically centralized but physically distributed control scheme for network management where the SDN controllers reside on top of physical hardware. The control plane maintains a global view of the network and performs the network management over all portions of the network, (i.e. core, backhaul, access, etc.). 

The space segment mainly consists of SDN-enabled GEO satellites which communicate with one another through laser link.  The communication between the SDN-enabled components in backhaul, RAN, and 5GC network relies on fiber, while the communication between the space and ground segments is done through the satellite gateways or RNs. We focus only the optimization of he ground segment, therefore we do not consider the mobility of the satellite layer and the inter-satellite links (ISLs) in our analysis. 

We model the terrestrial network as an undirected $\mathcal{T} = (\mathcal{V}, \mathcal{E})$ graph where $(u,v) \in \mathcal{E} $ if there is a link between the nodes $u,v \in \mathcal{V}$. Let $\mathcal{G}\subseteq \mathcal{V} $ be the set of all potential nodes for gateway placement and $\mathcal{K}\subseteq \mathcal{V}$ be the set of all potential switches for controller placement. We note that the sets $\mathcal{G}$ and $\mathcal{K}$ are not necessarily disjoint. A typical substrate node $v \in \mathcal{V}$ may satisfy one or more of the following statements: (i) A satellite gateway is placed at node $v$, (ii) node $v$ is an initial demand point of the terrestrial network, (iii) node $v$ hosts an SDN controller. 

Various objectives may be sought when solving the joint satellite gateway and SDN controller placement problem from reliability maximization, latency minimization, gateway and controller load balancing etc., depending on the design requirements. The optimal solution will introduce two sets of terrestrial locations $g_1, g_2, ..., g_m$, and $k_1, k_2, ..., k_n$ for gateway placement and controller placement, in respective order, together with the node-to-gateway, and node to controller assignment assignment.

Regarding the delay of the network we consider only propagation latency.
The propagation latency of a path in the network is the sum of the propagation delays over its constituting links. A GEO satellite is considered in the particular system model similar to \cite{kato17}, \cite{capac}, \cite{onlyrel}, and \cite{nar}. Let $d_{uv}$ represent the contribution of the terrestrial link $(u,v)$ to the propagation delay of a path which contains that link. The propagation delay from a gateway to the satellite is constant.

Furthermore, let us define the reliability $r^s_{ij}$ of each terrestrial-satellite path $P^s_{ij}$ from node $i \in \mathcal{V}$ to satellites $s$ which passes through gateway $j \in \mathcal{G}$ as in \cite{scgp}, where reliability of a path is modeled as the product of the reliability along its constituting components; i.e.

\begin{equation}
    r_{ju}^s =  (1 - P^{sg}_{e})\prod _{e \in P^s_{ju}} {(1 - P_e)} \prod_{u \in P^s_{ju} }{ (1 - P_v)} \quad \forall j \in \mathcal{G}, u \in \mathcal{V}
    \label{relst}
\end{equation}

In a similar fashion, the reliability of a controller-terrestrial node path is defined as the product of reliability along its constituting components as follows.

\begin{equation}
    r_{ku} = \prod _{e \in P_{ku}} {(1 - P_e)} \prod_{v \in P_{ku} }{ (1 - P_v)}
    \quad \forall k \in \mathcal{K}, u \in \mathcal{V}
\end{equation}

Table \ref{paramv} summarizes the notations reserved for variables and parameters frequently used in the paper.

\begin{table*}[ht]
\centering
\begin{center}
\scalebox{0.9}{
\begin{tabular}{|c|c|}
\hline
Variables & Description\\
\hline
$x_j$ & The binary decision variable of gateway placement at node $j$\\
$y_k$ & The binary decision variable of controller placement at node $k$\\
$w_{jv}$ & The binary decision variable for the assignment of node $v$ to gateway $j$ \\
$z_{jv}$ & The binary decision variable for the assignment of node $v$ to controller $j$ \\

\hline
Parameters & Description\\
\hline
$\mathcal{T}= (\mathcal{V},\mathcal{E})$ & Terrestrial network graph\\
$\mathcal{G}$ & The set of potential nodes for gateway placement \\
$\mathcal{K}$ & The set of potential nodes for controller placement \\
$m_{max}$ & The maximum number of permitted gateways for deployment\\
$k_{max}$ & The maximum number of permitted controllers for deployment\\
$d_{uv}$ & The propagation delay of the terrestrial link $(u,v)$\\
$r^s_{jv}$ & The reliability of the shortest path from node $v$ to satellite $s$ passing through gateway $j$\\
$r_{kv}$ & The reliability of the shortest path from node $v$ to SDN controller $k$\\
$P_e$ & Failure probability of edge $e$ from the terrestrial network \\
$P_v$ & Failure probability of node $v$ from the terrestrial network \\
$P^{sg}_e$ & Failure probability of gateway-satellite edge $e$ \\
$U(w, x, y, z)$ & Utility of the joint gateway and controller placement\\
$V(w, x, y, z)$ & Cost of the joint gateway and controller placement\\
\hline
\end{tabular}}
\caption{System model parameters and variables}
\label{paramv}
\end{center}
\end{table*}





%% file: formulations.tex
\section{Optimization Model \& Solution Approach}
\label{sec:problem}

We propose a two-step solution for the deployment of satellite gateways and the placement of the SDN controllers due to various sets of design requirements as hard constraints. In the first step, we model the gateway deployment problem as a MILP and determine the optimal placement policy for gateways, and then assuming the placement of gateways is computed and given as input to the controller placement phase, we formulate the SDN controller placement problem again as a MILP and determine the optimal controller placement policy. We use the sub (super)-modularity property and sub-modular optimization techniques that are described in detail in subsection \ref{sec:submdis}.

\subsection{ Problem MILP Formulation}
\label{milpform}
As explained in section \ref{sec:relatedwork}, within the framework of SAGIN, various objectives may be pursued by solving the problem at hand. One may want to minimize the cost of operation, over all number of deployed equipment, and the latency between the equipment and the demand points, or maximize the reliability of the chosen paths. In its most general form we wish to maximize a composite utility function which is a function of the placement of the gateways, the assignment of demand points to gateways, the placement of controllers, and the assignment of demand points to controllers. In other words, a total utility function is going to be maximized that is determined by the choice of the deployment, and the assignment policies. i.e. 

\begin{equation}
  Maximize \quad U(w,x,y,z) \label{objtot}
\end{equation}

with the following sets of decision variables: 

\begin{itemize}
    \item The set of binary assignment decision variables $\textbf{w}$ where $w_{ij} = 1$, if the traffic of node $j$ is assigned to the gateway placed at node $i$.
    \item The set of binary placement decision variables $\textbf{x}$ where $x_i = 1$, if a gateway is placed at node $i$. 
    \item The set of binary assignment decision variables $\textbf{z}$ where $z_{ij} = 1$, if the gateway placed at node $j$ is assigned to the controller placed at node $i$.
    \item The set of binary placement decision variables $\textbf{y}$ where $y_i = 1$, if a controller is placed at node $i$. 
\end{itemize}

and subject to various design requirements enumerated below. Each terrestrial node, whether it is a controller or not, has to be assigned to a gateway. i.e.,
\begin{equation}
    \sum_{j \in \mathcal{G}} w_{j v}=1 \quad \forall v \in \mathcal{V} \label{ass1}
\end{equation}
And the assignment of the terrestrial node $v$ to node $j$ is only valid if in the resulting placement policy, a gateway is located at node $j$.
\begin{equation}
 w_{j v} \leqslant x_{j} \quad \forall j \in \mathcal{G}, v \in \mathcal{V} \label{validg}   
\end{equation}



Each node has to be assigned to a controller. i.e.,

\begin{equation}
    \sum_{k \in \mathcal{K}} Z_{k v}=1 \quad \forall v \in \mathcal{V} \label{assC}
\end{equation}
And again this assignment has to be valid:
\begin{equation}
    Z_{k v} \leqslant y_{v} \quad \forall v \in \mathcal{V}, k \in \mathcal{K} \label{validc}
\end{equation}

In the case that the number of resources are limited, there is a maximum number of gateways and SDN controllers available. Therefore the number of facilities in the deployment policy should not exceed the maximum available: 

\begin{equation}
    \sum_{j \in \mathcal{G}} x_j\leqslant g_{max} \label{numgg}
\end{equation}

\begin{equation}
    \sum_{j \in \mathcal{G}} y_k\leqslant k_{max} \label{numcc}
\end{equation}
 where $g_{max}$ and $k_{max}$ are the maximum number of available gateways and SDN controllers to be deployed.



All constraints mentioned above, possibly together with other scenario-specific constraints may be embedded in the optimization problem depending on the design requirements and the a-priori knowledge of the network. The utility function $U$ can be a composite of multiple utility functions $U_i$. For instance we can consider the utility obtained by the minimization of the number of deployed gateways, and the utility obtained by maximizing the reliability of the controller-gateway paths. Similarly, objectives inspired by balancing the load on the gateways and/or the controllers can be taken into account, etc. . Corresponding to each utility function we can think of a cost function $V$ by simply negating the sign of the utility function. Therefore, the utility maximization problem can be also pursued as a cost minimization one.  Let us present the baseline MILP formulation for the joint satellite gateway deployment and SDN controller placement  as follows:

 In the case that the average reliability of the network is considered, we may have the corresponding utility function as: 
 
 \begin{equation}
     U^{rel}( w, x, y, z) = \frac{1}{|\mathcal{V}|}(\sum_{j \in \mathcal{G}}\sum_{v \in \mathcal{V}}{r^s_{jv}}{w_{jv}}+ \sum_{k \in \mathcal{K}}\sum_{v \in \mathcal{V}}{r_{kv}}{z_{kv}})
 \end{equation}
 
 where, the first term corresponds to the average reliability of node-to-gateway assignments, and the second term captures the same for the assignment to SDN controllers. 
 
 We note that within the context of SDN, the SDN controllers need to communicate information with one another regarding the state of the network in order to maintain a global view of the network. This is often denoted as the \textit{synchronization} cost in the literature. A poor configuration of the SDN-enabled network may result in huge excessive overhead contributing negatively to the cost of operation. It is therefore desirable to minimize the overhead corresponding to the synchronization between the SDN controllers. In \cite{tass}, the authors provide a realistic model for capturing the synchronization cost between the SDN controllers at edge which we adapt in this work with some modifications: 
 
 \begin{align}
     &V^C(w, x, y, z) = \sum_{m, n \in \mathcal{K}}{l^{(1)}_{mn}y_my_n} + \nonumber \\& \sum_{m, n \in \mathcal{K}}{y_my_nl^{(2)}_{mn}(\sum_{v \in \mathcal{V}}{z_{mv}})}+
     \sum_{m \in \mathcal{K}}{y_ml^{(3)}_m}
     \label{synch_tass}
 \end{align}
 
 where the first summation captures the constant communication cost between two controllers, the second term captures the cost that is dependant on the load of each controller, i.e. the number of users that are assigned to that controller. Finally, the third part of the synchronization cost captures the synchronization between the controllers on the ground segment to those in the space segment. In this paper, we assume that the constant cost between two ground controllers is mainly a function of the latency between them, i.e. $l^{(1)}_{mn} = d_{mn}$, the variable cost depends on the load of the controllers through a constant  $l^{(2)}_{mn} = l^{con}$, and the inter-segment cost depends on the distance of each controller to the gateway it is assigned to. In other words, the synchronization cost from equation \eqref{synch_tass}, can be written as: 
 
  \begin{align}
     & V_{sync}^C(w, x, y, z) = \sum_{m, n \in \mathcal{K}}{d_{mn}y_my_n} +  \nonumber \\&\sum_{m, n \in \mathcal{K}}{y_my_nl^{con}(\sum_{v \in \mathcal{V}}{z_{mv}})} +
     \sum_{m \in \mathcal{K}}{y_m(\sum_{j\in \mathcal{G}}{d_{jm}w_{jm}})}
     \label{synch_tass}
 \end{align}
 
 Therefore, the objective to minimize the average network latency, and the SDN controller synchronization overhead can be modeled as: 
 
 \begin{align}
     &V(w, x, y, z) = \sum_{j \in \mathcal{G}}{x_j} +\alpha\sum_{j \in \mathcal{G}}\sum_{v \in \mathcal{V}}{d_{jv}}{w_{jv}}+ \psi[\nonumber \\&\sum_{k \in \mathcal{K}}\sum_{v \in \mathcal{V}}{d_{kv}}{z_{kv}} + \beta V_{sync}^C(w, x, y, z)
] \end{align}
 where $\alpha$, $\beta$, and $\psi$,  are the corresponding factors to balance the emphasis of the objective function on different terms. 
 
 Therefore, we formulate the joint satellite gateway placement and SDN controller placement in an SDN-enabled SAGIN for reliability maximization as a MILP as follows:

 \begin{align}
 \text{Maximize} \quad &U^{rel}(w,x, y, z) \label{kol}\\
 \text{subject to:}
& \quad \eqref{ass1}, \eqref{validg}, \eqref{assC}, \eqref{validc}, \eqref{numgg}, \eqref{numcc}\\
& x_j \in \{0,1\}, \quad \forall j \in \mathcal{G}\\
& w_{jv} \in \{0,1\}, \quad \forall v \in \mathcal{V}, \quad \forall j \in \mathcal{G}\\
& z_{kv} \in \{0,1\}, \quad \forall k \in \mathcal{K}, \quad \forall v \in \mathcal{V}\\
& y_{k} \in \{0,1\}, \quad \forall k \in \mathcal{K} \label{lastdom}
\end{align}

 In order to formulate the problem for minimizing the average network latency and overhead, we have to apply some standard linearization methods as in \cite{tass}, to linearize the synchronization cost $V_{sync}^C(w, x, y, z)$. Namely, let us define a new set of binary variables $\left(\theta_{m n} \in\{0,1\}\right)$ to replace $\{{y_{m}y_{n}}\}$ term. 

Then we add the following linear constraints: 

\begin{align}
& \theta_{m n} \leq y_{m}, \quad \forall m, n \in \mathcal{K} \label{l11}\\
& \theta_{m n} \leq y_{n}, \quad \forall m, n \in \mathcal{K} \label{l12}\\
& \theta_{m n} \geq y_{m}+y_{n}-1, \quad \forall m, n \in \mathcal{K} \label{l13}
\end{align}

In a similar fashion, we introduce $\left(\phi_{m n v} \in\{0,1\}\right)$, and $\left(\mu_{m j} \in\{0,1\}\right)$ to replace $\{{y_{m}y_{n}z_{mv}}\}$, and, $\{y_m w_{jm}\}$ in respective order, and add the following constraints:

\begin{align}
& \phi_{m n v} \leq \theta_{mn}, \quad \forall m, n \in \mathcal{K}, v \in \mathcal{V} \label{l21}\\
& \phi_{m n v} \leq y_{m}, \quad \forall m, n \in \mathcal{K}, v \in \mathcal{V} \label{l22}\\
& \phi_{m n v} \geq y_{m}+\theta_{mn}-1, \quad \forall m, n \in \mathcal{K}, v \in \mathcal{V} \label{l23}
\end{align}

\begin{align}
& \mu_{m j} \leq y_{m}, \quad \forall m \in \mathcal{K}, j \in \mathcal{G} \label{l31}\\
& \mu_{m j} \leq w_{jm}, \quad \forall m \in \mathcal{K},j \in \mathcal{G} \label{l32}\\
& \mu_{m j} \geq y_{m}+w_{jm}-1, \quad \forall m \in \mathcal{K}, j \in \mathcal{G} \label{l33}
\end{align}

towards achieving the following MILP formulation.

 \begin{align}
 \text{Mainimize} \quad &V^{lat+ovrh.}(w,x, y, z) \label{kol}\\
 \text{subject to:}
& \quad \eqref{ass1}, \eqref{validg}, \eqref{assC}, \eqref{validc}, \eqref{l11},  \eqref{l12}, \eqref{l13}, \eqref{l21}, \nonumber \\&\eqref{l22}, \eqref{l23}, \eqref{l31}, \eqref{l32}, \eqref{l33}\nonumber\\
& x_j \in \{0,1\}, \quad \forall j \in \mathcal{G}\\
& w_{jv} \in \{0,1\}, \quad \forall v \in \mathcal{V}, \quad \forall j \in \mathcal{G}\\
& z_{kv} \in \{0,1\}, \quad \forall k \in \mathcal{K}, \quad \forall v \in \mathcal{V}\\
& y_{k} \in \{0,1\}, \quad \forall k \in \mathcal{K} \label{lastdom}
\end{align}
 
 \begin{figure*}[h]
\begin{center}
\begin{minipage}[h]{0.425\textwidth}
\includegraphics[width=1\linewidth]{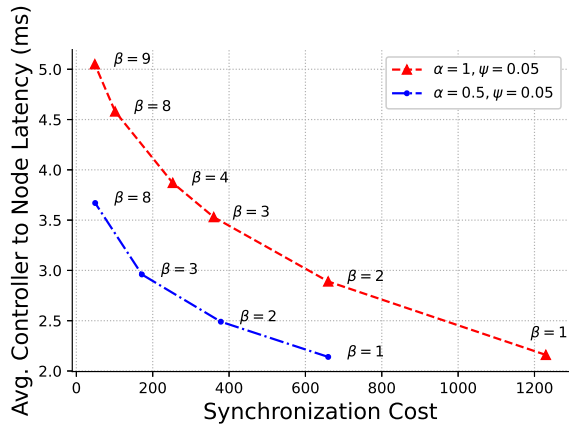}
\caption{ Trade-off between the synchronization cost and average controller-to-node latency}
\label{fig:first_tradeoff}
\end{minipage}
\end{center}
\end{figure*}

 Fig. \ref{fig:first_tradeoff}, illustrates the impact of parameters $\alpha$, and $\beta$, on the solution of the problem, while keeping $\psi$ constant, where the trade-off between the synchronization cost and the average controller-to-node latency is depicted for two different cases of $\alpha$. When $\alpha = 0.5$ is set, by increasing the value of $\beta$, more emphasis is put on minimizing the synchronization cost compared to the average controller-to-node latency. A similar trend is observed when setting $\alpha = 1$, but with the curve shifted upwards; since increasing the value of $\alpha$, will put more emphasis on optimizing the gateway placement policy.

 The joint satellite gateway and SDN controller placement problem is an instance of the facility location problem that is known to be $\mathcal{NP}$-hard. Commercial MILP solvers such as CPLEX can be used directly to solve such models for small instances using well-established methods such as branch\& bound, branch\& cut, etc., however due to the large number of variables, the problem will soon become intractable for large instances, and such methods will tend to be ineffective. Therefore, to overcome this issue, it is crucial to come up with approximation methods that scale well with the size of the network and generate acceptable solutions both in terms of time results accuracy and time complexity. 
 
 In the next subsection we give a short review of submodular optimization and then will justify that while specifying the utility/cost functions, our model can suitably fit into the submodular optimization framework subject to constraints of known types. We then apply two interesting algorithms from the submodular optimization literature to efficiently solve our problems.


\subsection{Sub (Super)-modular Optimization Methods}
\label{sec:submdis}

Let us start with the definition of submodular functions. 

\begin{defn}\textbf{Submodular Functions.} Let finite set  $G$ of elements be the ground set. Then A function $f: 2^{G} \rightarrow \mathbb{R}$ over the ground set is said to be submodular if for all subsets $A, B \subseteq G$, it holds that $$f(A)+f(B) \geq f(A \cup B)+f(A \cap B)$$

Equivalently, $f$ is said to be submodular if for all subsets $A, B \subseteq G$, 
with $A \subseteq B$ and every element $i \in G \backslash B$ it holds that:
$$
f(A \cup\{i\})-f(A) \geq f(B \cup\{i\})-f(B)
$$
\end{defn}

This intuitively means that for a submodular set function, adding an element to a subset will result in diminishing return with increasing the subset size. 
We also note that if for all subsets $A, B \subseteq G$, 
with $A \subseteq B$ it holds that $f(A) \leq f(B)$, then $f$ is called a \textit{monotone submodular function}. 

Moreover, it is worth noting that, if $f, g$ are submodular functions then $[f+g],[k f, k>0],[-f]$, are submodular, submodular, and supermodular in order.

Given the diminishing return property of the submodular functions, many utility functions can suitably fit in this class. Therefore, motivated by the natural application of submodularity property in real-world scenarios such as the welfare maximization, social networks, information gathering, feature selection, etc., optimization problems involving submodular/supermodular functions have developed a lot of interest among the research community. Especially, over the past decade, a lot of interesting methods have been proposed for approximately solving the submodular/supermodular optimization problems subject to a variety of constraints while providing acceptable optimality bounds. Within the SDN research community, several papers have used submodular optimization to model and solve multiple resource allocation problems \cite{cont}\cite{cont2}\cite{tass}.
Moreover, in \cite{tass2}, a very interesting taxonomy of such problems in the mobile edge computing (MEC) framework, along with an insightful discussion on their submodularity property is provided. 

In this paper we show that the utility functions corresponding to the objectives of our interest are submodular or the corresponding cost functions can be modeled as supermodular functions. Then we use two interesting algorithms from the submodular optimization literature to efficiently come up with approximate solutions. Let us first state two theorems that we will use in the subsequent sections. 

\begin{thm}[\cite{card}]
There exists a $(1 - 1/e -\varepsilon)$-approximation algorithm for maximizing a monotone non-negative submodular function with cardinality constraints.  The algorithm has a $\mathcal{O} (\frac{n}{\varepsilon} \log{\frac{n}{\varepsilon}})$ time complexity. \label{moncard}
\end{thm}

\begin{thm}[\cite{focs}]
 There exists a $1/2$-approximation randomized greedy algorithm for maximizing a non-negative submodular function, which runs in linear time. 
 \label{yedovvom}
\end{thm}






\subsection{Problem Decomposition}
The joint placement problem specified in equations \eqref{kol}-\eqref{lastdom} contains variables both corresponding to the placement of gateways and the placement of controllers. This will render a rather complicated problem with nonlinear constraints due to the unknown location of both gateways and controllers. 
Therefore, to simplify the problem we adopt a sequential two-step approach. Since the location of the gateways has great impact on both latency and reliability requirements, we first determine the optimal placement of the gateways and then choose the best controller placement policy accordingly. Hence, we can decompose the utility function in \eqref{kol} and write:
$$ U(w,x,y,z) = U_g(w,x) + U^{wx}_c(y,z)$$ 

where the first term is concerned only with the placement of the gateways and the second term corresponds to the utility obtained by placing the SDN controllers given that the location and the assignment of gateways is determined. Similarly, if a cost function is to be minimized we obtain:

$$ V(w,x,y,z) = V_g(w,x) + V^{wx}_c(y,z)$$

Under this decomposition we're dealing with two sub-problems: $(i)$ \textit{Satellite Gateway Deployment Problem}, and $(ii)$ \textit{SDN Controller Placement Problem}. In the next subsections we address these two problems sequentially.


\subsection{Optimal Gateway Placement}

Following the discussion in section \ref{milpform}, multiple objectives may be sought when deciding the optimal gateway placement policy. For instance, the network provider may intend to minimize the average network latency while at the same time minimizing the number of gateways deployed. 
Thus, an optimal gateway placement policy is one that minimizes the aggregated cost function. 


Given the above discussion we formulate the gateway placement problem for minimizing the aggregated cost (delay-oriented) with the objective function: 

\begin{equation}
    V_g(w,x) = V^1_g(w,x) + \alpha V^2_g(w,x)
    \label{v_delay}
\end{equation}
where, 
\begin{equation}
    V^1_g(w,x) = \sum_{j \in \mathcal{G}}{x_j} \label{first}
\end{equation}
\begin{equation}
    V^2_g(w,x) = \sum_{v \in \mathcal{V}}\sum_{j  \in \mathcal{G}} d_{jv}w_{jv} \label{second}
\end{equation}
The corresponding MILP model is as follows:

\begin{align}
 \text{Minimize} \quad &V_g(w,x) \label{ooo}\\
 \text{subject to:}
& \quad \eqref{ass1}, \eqref{validg}\\
& x_j \in \{0,1\}, \quad \forall j \in \mathcal{G}\\
& w_{jv} \in \{0,1\}, \quad \forall v \in \mathcal{V}, \quad \forall j \in \mathcal{G} \label{ccc}
\end{align}

Next we argue by the following theorem that if the placement of gateways is known, then the optimal assignment policy can be uniquely determined and then use this observation to show that the cost function $V_g(w,x)$ is supermodular. 
\begin{thm}
Given the placement of the gateways $x$, the optimal assignment policy $w^*$ can be uniquely determined; i.e. there exists a deterministic function $g: G\rightarrow W$ for which $w^* = g(x)$. \label{thm1}
\end{thm}
\textbf{Proof.} Once the placement of gateways is determined, $V^1_g(w,x)$ is already decided. Let $\mathcal{X} \subseteq \mathcal{G}$ be the set of nodes where a gateway is placed. In other words, 
\begin{equation}
    \mathcal{X} = \{j \in \mathcal{G}: x_j =1\}
\end{equation}

To minimize the objective function \eqref{v_delay} the optimal assignment policy is as follows: 
\begin{equation}
 w^*_{jv} =  \mathds{1}_{\{j = \arg \min_{j \in \mathcal{X}} d_{jv}\}} , \quad \forall j \in G, v \in \mathcal{V} \label{asspolicy}.   
\end{equation}
which implies $w^* = g(x)$. {$\blacksquare$}

\begin{thm}
The cost function $V_g(w, x)$ is supermodular. \label{gate_min}
\end{thm}
\textbf{Proof.} Let $\mathcal{A},\mathcal{B} \in \mathcal{G}$ be two arbitrary sets of locations corresponding to two separate gateway placement policies such that  $\mathcal{A}\subseteq \mathcal{B}$. Let us update the policies $\mathcal{A}$, and $\mathcal{B}$ by adding a new gateway at location $g \in \mathcal{G}\setminus \mathcal{B}$. For any set function $f$, let $R^{\mathcal{A}}_{g} (f)$ be the amount of change in the cost function by adding the new gateway $g$ to policy $\mathcal{A}$. $R^{\mathcal{B}}_{g} (f)$  can be defined in a similar fashion. i.e.
$$ R^{\mathcal{A}}_{g} (f) = f (\mathcal{A}\cup \{g\}) - f (\mathcal{A}) $$

By the definition of the supermodular functions, a function $f$ is supermodular if $R^{\mathcal{A}}_{g}$ $\leq$ $R^{\mathcal{B}}_{g}$. $V^1_g(w, x)$ only depends on the placement policy, and the marginal return of adding any new gateway is constant. Thus, $R^{\mathcal{A}}_{g}$ = $R^{\mathcal{B}}_{g}$ and $V^1_g(w, x)$ is modular and therefore supermodular. 

In order to prove the supermodularity of $V^2_g(w, x)$, let $    j^\mathcal{A}_v =: \arg\min_{j \in \mathcal{A}}{d_{jv}}$ be the location of the gateway to which node $v\in \mathcal{V}$ is assigned under gateway placement policy $\mathcal{A}$. The definition for $j^\mathcal{B}_v$ follows in a similar fashion. Moreover, let us consider the set of all nodes that can contribute to reducing the cost of gateway placement by switching the recently introduced gateway:

\begin{equation}
    \sigma(\mathcal{V}) = \{v \in \mathcal{V}: d_{gv} \leq \min_{j \in \mathcal{B}}{d_{jv}}\}
\end{equation}

Then we can write: 



\begin{align}
    &R^{\mathcal{B}}_{g} - R^{\mathcal{A}}_{g} = \sum_{v \in \mathcal{V}}{\min(0, (d_{gv} - d_{j^{\mathcal{B}}_vv}))}\nonumber \\& - \sum_{v \in \mathcal{V}}{\min(0, (d_{gv} - d_{j^{\mathcal{A}}_vv}))}\label{mmmmm}\\
     &\geq \sum_{v \in \sigma(\mathcal{V})}{ (d_{gv} - d_{j^{\mathcal{B}}_vv})} - \sum_{v \in \sigma(\mathcal{V})}{ (d_{gv} - d_{j^{\mathcal{A}}_vv})} \label{vasat} \\
     &=\sum_{v \in \sigma(\mathcal{V})}{(d_{j^{\mathcal{B}}_vv} - d_{j^{\mathcal{A}}_vv})} \geq 0.
\end{align}

where each node $v$ contributes to the cost reduction if and only if it can get closer to its nearest gateway by switching to the newly added gateway $g$. The lower bound in equation \ref{vasat}, comes from the fact that it only includes a subset of such nodes for policy $\mathcal{A}$; therefore potential cost savings may be left out of equation \ref{vasat}. {$\blacksquare$}


\vspace{2mm}
As explained in section \ref{sec:submdis}, the literature on the approximation algorithms for optimizing submodular functions is very rich where most of the interesting results are derived for the case of non-negative functions. Also, it is worth noting that minimizing a supermodular function may be approached by first casting the optimization problem as a non-negative submodular maximization and then utilizing the corresponding approximation algorithms. 

With such an approach, the authors in \cite{tass} have previously used the algorithm mentioned in theorem \ref{yedovvom}, for SDN controller placement at edge. In a similar fashion, let $\Bar{V}_g$ be the maximum of equation \eqref{v_delay}. Thus, minimizing \eqref{v_delay} is tantamount to maximizing 

\begin{equation}
    \hat{V}_g(\mathcal{X}) = \Bar{V}_g-(V^1_g(\mathcal{X}) + V^2_g(\mathcal{X})) 
    \label{v_delay2}
\end{equation}
that is a non-negative submodular function, in which the dependence on $w$ is dropped according to theorem \ref{thm1}.
 Algorithm \ref{alg1} summarizes  how the $(1/2)$-approximation procedure occurs.  
 
\begin{algorithm}
 \caption{(1/2)-approximation greedy algorithm}
 \label{alg1}
 \begin{algorithmic}[1]
  \renewcommand{\algorithmicrequire}{\textbf{Input:}}
  \renewcommand{\algorithmicensure}{\textbf{Output:}}
  \REQUIRE $\hat{v}_g: 2^\mathcal{G} \rightarrow \mathds{R}_+$
  \ENSURE $(\Bar{X}, \hat{v}_g(\Bar{X}))$
\STATE Initialize \quad $\underaccent{\bar}X = \emptyset , \quad \Bar{X} = \{1\}^{|\mathcal{G}|} $
 \STATE \textbf{for $j \in \mathcal{G}:  $}
 \STATE \quad $\Bar{\Delta} = max(\hat{V}_g(\Bar{X}) - \hat{V}_g(\Bar{X}\setminus \{i\}), 0)$
 \STATE \quad $\underaccent{\bar}\Delta = max( \hat{V}_g(\underaccent{\bar}{X}) - \hat{V}_g(\underaccent{\bar}{X}\cup \{i\}), 0)$
 \STATE \quad \textbf{Set} $\quad  \underaccent{\bar}{X} = \underaccent{\bar}{X}\cup \{i\}$ with probability $\frac{\underaccent{\bar}\Delta}{(\underaccent{\bar}\Delta + \Bar{\Delta})}$
 \STATE \quad otherwise
 \STATE \quad \textbf{Set} $\quad \Bar{X} = \Bar{X}\setminus \{i\}$ 
 \STATE \textbf{end}
 \RETURN $(\Bar{X}, \hat{v}_g(\Bar{X}))$
 \end{algorithmic} 
 \end{algorithm}
 
 The algorithm starts by taking two extreme cases and then  decides on the placement of a gateway at each location in an iterative fashion. Before $i$th iteration begins, a gateway is present in $i$th location by the policy $\Bar{X}$ and absent by the policy $\underaccent{\bar}X$. The algorithm computes the contribution of the inclusion/exclusion of a gateway in $i$th location, and makes a randomized choice accordingly. After iteration $i$ ends both the policies agree on the inclusion/exclusion of a gateway at location $i$. Hence, when the execution of the algorithm finishes the two policies will be the same.

It is important to note that in realistic cases the number of facilities and the amount of resources are limited. 
Therefore, it makes sense to formulate the gateway placement optimization problem under an additional \textit{cardinality} constraint. Let us consider the reliability-oriented utility function. i.e.

\begin{equation}
    U_g(w,x) = \sum_{v \in \mathcal{V}}\sum_{j \in \mathcal{G}} r^s_{jv}w_{jv}
\end{equation}


Therefore, the MILP model will be as follows: 

\begin{align}
 \text{Maximize} \quad &U_g(w,x)\\
 \text{subject to:}
& \quad \eqref{ass1}, \eqref{validg}, \eqref{numgg}\\
& x_j \in \{0,1\}, \quad \forall j \in \mathcal{G}\\
& w_{jv} \in \{0,1\}, \quad \forall v \in \mathcal{V}, \quad \forall j \in \mathcal{G}
\end{align}

 In the nest theorem, we claim that $U_g(w,x)$ is a non-negative \textit{submodular} function that is monotone as well. 

 \begin{thm}
 $U_g(w, x)$ is a monotone submodular function. \label{max_gate}
\end{thm}
\textbf{Proof.} Similar to the proof of theorem \ref{gate_min}, consider any $\mathcal{A},\mathcal{B} \in \mathcal{G}$ as two arbitrary sets of locations corresponding to two separate gateway placement policies such that  $\mathcal{A}\subseteq \mathcal{B}$ and suppose a new gateway is going to be deployed at location $g \in \mathcal{G}$. To see that  $U_g(w, x)$ is an increasingly monotone function, observe that:
\begin{equation}
    \forall v \in \mathcal{V}:  r^s_{j^\mathcal{B}_vv} \geq r^s_{j^\mathcal{A}_vv}
\end{equation}
 
 where $j^\mathcal{A}_v$, and $j^\mathcal{B}_v$ maximize the reliability of assignment for node $v$. The assertion follows by summing the $LHS$ and the $RHS$ of the last inequality, over $v \in \mathcal{V}$. 
 
 To show the submodularity of  $U_g(w, x)$, we need to compare $R^{\mathcal{B}}_{g}$ and $R^{\mathcal{B}}_{g}$. With a similar logic define:
 
 \begin{equation}
    \sigma(\mathcal{V}) = \{v \in \mathcal{V}: r^s_{gv} \geq \max_{j \in \mathcal{B}}{r^s_{jv}}\}
\end{equation}

 It holds that: 
 
 \begin{align}
     &R^{\mathcal{B}}_{g} - R^{\mathcal{A}}_{g} \leq \sum_{v \in \sigma(\mathcal{V})}{ (r^s_{gv} - r^s_{j^{\mathcal{B}}_vv})} - \sum_{v \in \sigma(\mathcal{V})}{ (r^s_{gv} - r^s_{j^{\mathcal{A}}_vv})} = \nonumber\\ &\sum_{v \in \sigma(\mathcal{V})}{(r^s_{j^{\mathcal{A}}_vv}-r^s_{j^{\mathcal{B}}_vv})} \label{vassssssat} \leq 0 
 \end{align}
 
 Therefore, by the definition of submodularity,  $U_g(w, x)$ is submodular. {$\blacksquare$}
\vspace{4mm}

To approximately solve this problem, we use theorem \ref{moncard}. Algorithm \ref{alg2} outlines the corresponding procedure based on theorem \ref{moncard}.

 \begin{algorithm}
 \caption{$(1 - 1/e -\varepsilon)$-approximation algorithm}
 \label{alg2}
 \begin{algorithmic}[1]
  \renewcommand{\algorithmicrequire}{\textbf{Input:}}
  \renewcommand{\algorithmicensure}{\textbf{Output:}}
 \REQUIRE ${U}_g: 2^\mathcal{G} \rightarrow \mathds{R}_+, s \in \{1,...,|\mathcal{G}|\}$, $0<\varepsilon<1$
 \ENSURE $({X}, {U}_g({X}))$
\STATE \textbf{Initialize}: $X \leftarrow \emptyset$, and $d \leftarrow \max_{j \in \mathcal{G}}{U_g(\{j\})}$
\STATE $\textbf { for }\left(w=d ; w \geq \frac{\epsilon}{n} d ; w \leftarrow w(1-\epsilon)\right)$
\STATE \quad \textbf{for} $j \in \mathcal{G}$ 
\STATE \quad \quad \textbf{if} $|X \cup\{j\}| \leq s$ and $U_g({X \cup\{j\}}) -  U_g({X}) \geq w$ 
\STATE \quad \quad \quad $X \leftarrow X \cup\{j\}$
\STATE \quad \quad \textbf{endif}
\STATE \quad \textbf{end}
\STATE \textbf{end}
 \RETURN $({X}, U_g({X}))$
 \end{algorithmic} 
 \end{algorithm}
 
 The algorithm starts by finding the most beneficial single gateway placement, i.e. a single gateway that provides the highest average network reliability among all others. Then according to this maximum contribution, it iteratively picks a value for minimum contribution and adds all the gateways that can be more beneficial than that, all in a single iteration. In other words, at each iteration all those locations that can pass the minimum contribution level, are eligible for placement of a gateway. The minimum contribution decreases in each iteration according to a pre-defined step-size and therefore the criterion for eligibility becomes less strict over time. If at any step before the end of the loop, the number of the deployed gateways reaches that maximum allowed $s$, the algorithm terminates immediately.

\subsection{Optimal Controller Placement}
Once the placement of the gateways is determined, we need to choose the optimal controller placement policy accordingly. Since controllers are expensive resources, we wish to deploy as few controllers as possible. Furthermore, the paths from gateways to controllers must be reliable enough, in order to prevent disconnections between the control and the data plane. Finally, since the SDN controllers need to maintain a global view of the network, they will have to synchronize by communicating information regarding the global state of network. As explained in section \ref{milpform}, regardless of how this is achieved, the rate of communication needed between any pair of controllers is proportional to the load of each controller, in addition to the constant rate. Therefore, the cost of this communication needs to be taken into account which can be a function of the location of the controllers, the latency between them, etc.; The objective function of the SDN controller placement problem (latency/overhead oriented), consists of four parts:

 
 

\begin{align}
V^{wx}_c(y,z) =  & V^{wx}_{c_1}(y,z) + \beta [V^{wx}_{c_2}(y,z)+  V^{wx}_{c_3}(y,z) + \nonumber\\& V^{wx}_{c_4}(y,z) ]  \label{conob}    
\end{align}

where,

\begin{align}
&V^{wx}_{c_1}(y,z) = \sum_{k \in \mathcal{K}}\sum_{v \in \mathcal{V}}{d_{kv}}{z_{kv}} \\
&V^{wx}_{c_2}(y,z) = \sum_{m, n \in \mathcal{K}}{d_{mn}y_my_n} \\
&V^{wx}_{c_3}(y,z) = \sum_{m, n \in \mathcal{K}}{y_my_nl^{con}(\sum_{v \in \mathcal{V}}{z_{mv}})} \\
& V^{wx}_{c_4}(y,z) = \sum_{k \in \mathcal{K}}\sum_{j \in \mathcal{X}}{d_{jm}y_m}
\end{align}




Therefore, considering the analysis in section \ref{milpform} regarding linearization of the objective function, we obtain the following MILP model for the SDN controller placement problem: 
\begin{align}
 \text{Minimize} & \sum_{k \in \mathcal{K}}\sum_{v \in \mathcal{V}}{d_{kv}}{z_{kv}} + \beta [ \sum_{m, n \in \mathcal{K}}{d_{mn}\theta_{mn}} + \nonumber\\ 
 &\sum_{m ,n \in \mathcal{K}}\sum_{v \in \mathcal{V}}{l^{con}\phi_{mnv}} + \sum_{k \in \mathcal{K}}\sum_{j \in \mathcal{X}}{d_{jm}y_m}] \\
 \text{subject to:}
& \quad \eqref{assC}, \eqref{validc}, \eqref{l11}-\eqref{l23} \\
& z_k \in \{0,1\}, \quad \forall k \in \mathcal{K}\\
& y_{jk} \in \{0,1\}, \quad \forall k \in \mathcal{K}, \quad \forall j \in \mathcal{X}
\end{align}

In the next theorem, we show that the function $V^{wx}_c(y,z)$ is sub-modular, and accordingly the $(1/2)$-approximation method can be used to approximately minimize it. 

\begin{thm}
The cost function $V^{wx}_c(y,z)$ is supermodular.
\end{thm}
\textbf{Proof.} We will first show the supermodularity of each of the $4$ terms. As usual, consider any $\mathcal{A},\mathcal{B} \in \mathcal{K}$ such that $\mathcal{A}\subseteq \mathcal{B}$ to be two sets of nodes selected to host SDN controllers by two different controller placement policies, where $k^\mathcal{A}_v =: \arg\min_{k \in \mathcal{A}}{d_{kv}}$ for all $v \in \mathcal{V}$, is the controller that node $v$ is assigned to, by the assignment policy implied by the placement $\mathcal{A}$.  $k^\mathcal{B}_v$ can be similarly defined for set $\mathcal{B}$. 

By introducing a new SDN controller to both of the policies at any arbitrary location $\kappa \in \mathcal{K} \setminus \mathcal{B}$ and substituting the parameters $z_{kv}$, $d_{kv}$, $k^{\mathcal{A}}_{v}$, and $k^{\mathcal{B}}_{v}$ with their counterparts in the statement of the proof of theorem \ref{gate_min}, it will follow that $R^{\mathcal{B}}_{\kappa} \geq R^{\mathcal{A}}_{\kappa}$; hence $V^{wx}_{c_1}(y,z)$ is a supermodular function.

The supermodularity of the function $V^{wx}_{c_2}(y,z)$ is immediate, since once adding a new controller to sets $\mathcal{A}$, and, $\mathcal{B}$, since the set $\mathcal{B}$ is bigger, the new controller has to communicate with more controllers compared to when it is added to set $\mathcal{A}$; therefore $R^{\mathcal{B}}_{\kappa} \geq R^{\mathcal{A}}_{\kappa}$.

The function $V^{wx}_{c_3}(y,z)$, is modular. To see this we can change the order of summations in $V^{wx}_{c_3}(y,z)$ and write: 

\begin{equation}
    V^{wx}_{c_3}(y,z) = \sum_{v \in \mathcal{V}} {(\sum_{k \in \mathcal{K}: z_{kv}=0}l^{con}y_k)}
\end{equation}
Now, observe that adding a new element to sets $\mathcal{A}$, and, $\mathcal{B}$, will only contribute $l^{con}$ for each $v \in \mathcal{V}$, regardless of the size of the set. Therefore $R^{\mathcal{B}}_{\kappa} - R^{\mathcal{A}}_{\kappa} =0$, and the function is modular. 

Similarly,  $V^{wx}_{c_3}(y,z)$ is modular, since the marginal return by adding a new controller is proportional to the distance of that controller to its closest gateway, regardless of the cardinality of the set of the controllers. 

Therefore, given that modularity is a special case of supermodularity,  $V^{wx}_c(y,z)$ as the sum of supermodular and modular functions is supermodular and the assertion follows. {$\blacksquare$}

\vspace{4mm}
Given that in the SDN-enabled hybrid network, the SDN controllers perform the task of network management, any failures in the nodes or links of the network my disconnect the data and the control plane and block the way of the instructions from controllers to the SDN switches which in turn will result in the performance degradation of the hybrid network.  Therefore, it makes sense to formulate the reliable controller placement problem with the objective of maximizing the average reliability of the control paths. i.e. 

\begin{equation}
    U^{wx}_c(y, z)  = \sum_{k \in \mathcal{K}}\sum_{v \in \mathcal{V}}{r_{kv}}{z_{kv}}
\end{equation}

The MILP model is as follows:

 \begin{align}
 \text{Maximize} \quad & U^{wx}_c(y, z) \label{kol}\\
 \text{subject to:}
& \quad \eqref{assC}, \eqref{validc}, \eqref{numcc}\\
& z_{kv} \in \{0,1\}, \quad \forall k \in \mathcal{K}, \quad \forall v \in \mathcal{V}\\
& y_{k} \in \{0,1\}, \quad \forall k \in \mathcal{K} \label{lastdom}
\end{align}

Next we show that the function $U^{wx}_c(y, z) $ is submodular and hence the algorithm mentioned in theorem \ref{moncard} can be applied for approximately solving the corresponding MILP model. 

\begin{thm}
The function $U^{wx}_c(y, z)$ is monotone and submodular.
\end{thm}
\textbf{Proof.} The proof is exactly similar to that of theorem \ref{max_gate} when $z_{kv}$, $R^{\mathcal{A}}_{\kappa}$, $R^{\mathcal{B}}_{\kappa}$, $r_{kv}$, $k^{\mathcal{A}}_{v}$, and $k^{\mathcal{B}}_{v}$  replace their counterparts in the statement of the proof.

%% file: evaluation.tex
\section{Performance Evaluation}
\label{sec:evaluation}
In this section we evaluate the performance of the proposed gateway and controller placement methods. We first provide an explanation on the evaluation setup and the simulation environment and then provide the corresponding results for each method. 

\subsection{ Evaluation Setup \& Metrics}
Similar to \cite{nar}, for our simulations, we utilize multiple real network topologies publicly available at the Internet Topology Zoo \cite{zoo}. In addition to the previous topologies used in \cite{nar} we use $4$ new networks to be able to compare the performance of our algorithms to that of the literature. The complete list of topologies that we have considered is listed in table \ref{topo}. The lengths of network links are extracted from the Topology Zoo, based on which the corresponding latency value for each link is computed in a similar fashion to \cite{nar}. 
To compute the shortest paths between each pair of network nodes, we adopted an implementation of the Yen's algorithm \cite{yen}. To facilitate the comparison of the algorithms with literature we apply the same approach as in \cite{onlyrel} to compute the failure probability for each network component. We randomly generate the failure probabilities for terrestrial nodes, terrestrial links, and the satellite link, in $4$ different settings. Table \ref{failure} lists the range of failure probabilities in different cases for each network component. For solving the MILP models we use CPLEX commercial solver, and conduct all the experiments on an Intel Xeon processor at 3.5 GHz and 16 GB of main memory. Unless otherwise stated, each experiment is repeated $100$ times and the results are averaged over all runs. 

\vspace{2mm}

\begin{table}[t]
\centering
\caption{Network Topology Settings}
\begin{tabular}{||c c c||} 
 \hline
 Topology & Nodes & Links  \\ 
 \hline\hline
 Nsfnet & 13 & 15  \\
 Sinet & 13 & 18  \\ 
 Ans & 18 & 25  \\
 Aarnet & 19 & 24\\
 Agis & 25 & 32  \\
 Digex & 31 & 35  \\
 Chinanet & 42 & 86\\
 Bell Canada & 48 & 64 \\ 
 Tinet & 53 & 89\\
 \hline
\end{tabular}
\label{topo}

\vspace{4mm}
\caption{Failure Probability Settings}
\begin{tabular}{|c|c|c|c|}
\hline & $P_{v}$ terrestrial nodes & $P_{e}$ terrestrial links & $P_{e_{s} g}$ satellite links \\
\hline Case 1 & {[0,0.05]} & {[0,0.02]} & {[0,0.02]} \\
\hline Case 2 & {[0,0.06]} & {[0,0.04]} & {[0,0.03]} \\
\hline Case 3 & {[0,0.07]} & {[0,0.06]} & {[0,0.04]} \\
\hline Case 4 & {[0,0.08]} & {[0,0.08]} & {[0,0.05]} \\
\hline
\end{tabular}
\label{failure}
\end{table}

        


\vspace{4mm}
In order to evaluate the performance of the proposed methods we use multiple metrics as follows:

\begin{itemize}

    \item \textbf{Utility Function Value} is the overall value of the utility function provided by the solution of the JGCP problem. 
    
    \item \textbf{Cost Function Value} is the overall value of the cost function provided by the solution of the JGCP problem. 
    
    \item \textbf{Average node-to-facility Latency} is the average shortest-path node-to-gateway or node-to-controller latency experienced by the terrestrial network. 
    
    \item \textbf{Average Network node-to-facility Reliability} is the average reliability of the shortest path between each terrestrial node and and a gateway or a controller.
    
    \item \textbf{Number of Deployed Facilities} is the total number of satellite gateways or controllers that is deployed on the terrestrial network as a result of the JGCP.
    
    \item \textbf{Solver Run-time} is the amount of time it takes for the solver to generate the solution to the JGCP problem. 
\end{itemize}

\subsection{Numerical Results}
We conduct $4$ experiments to evaluate the performance of our algorithms for the different problems discussed in section \ref{sec:problem}. The first two experiments correspond to the gateway placement problem. 

In \textit{Exp. A}, the latency-oriented gateway placement problem is considered with the objective of jointly minimizing the number of deployed gateways and the average network-to-gateway latency. We discuss the trade-off between the number of deployed gateways and the resulting average network-to-gateway latency.

\textit{Exp. B} aims at maximizing the average reliability of node-to-controller paths, while the number of gateways to be deployed are limited not to exceed a given maximum. We compare our results with the algorithms mentioned in \cite{onlyrel}, and also show how our method works under different settings for network failure probability. 

\begin{figure*}[t]
\begin{center}

\begin{minipage}[h]{0.45\textwidth}
\includegraphics[width=1\linewidth]{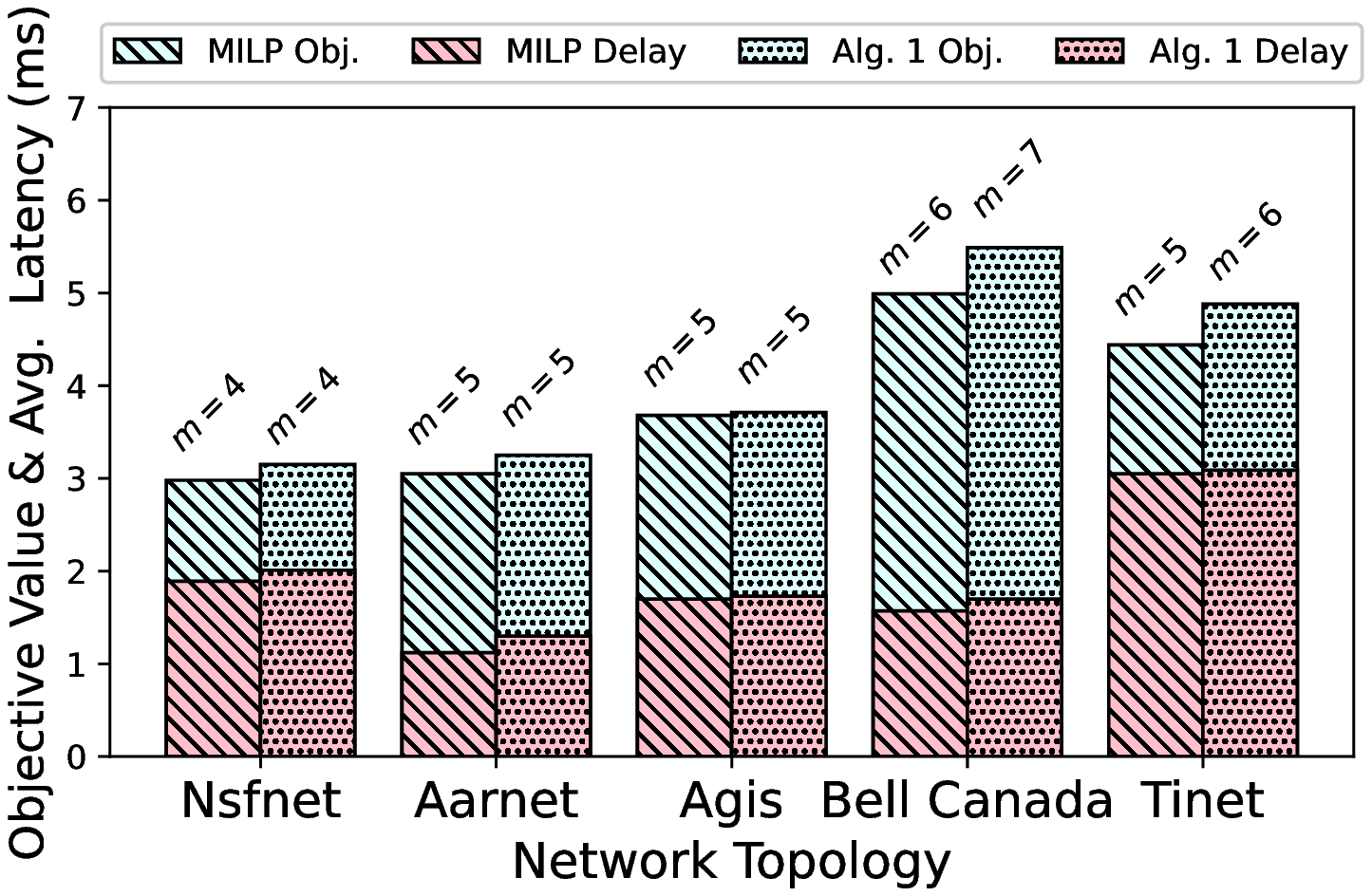}
\caption{ Exp A. Objective function \& average node-to-gateway latency}
\label{fig:lat_gw_comp}
\end{minipage}
\hspace{1mm}
\begin{minipage}[h]{0.425\textwidth}
\includegraphics[width=1\linewidth]{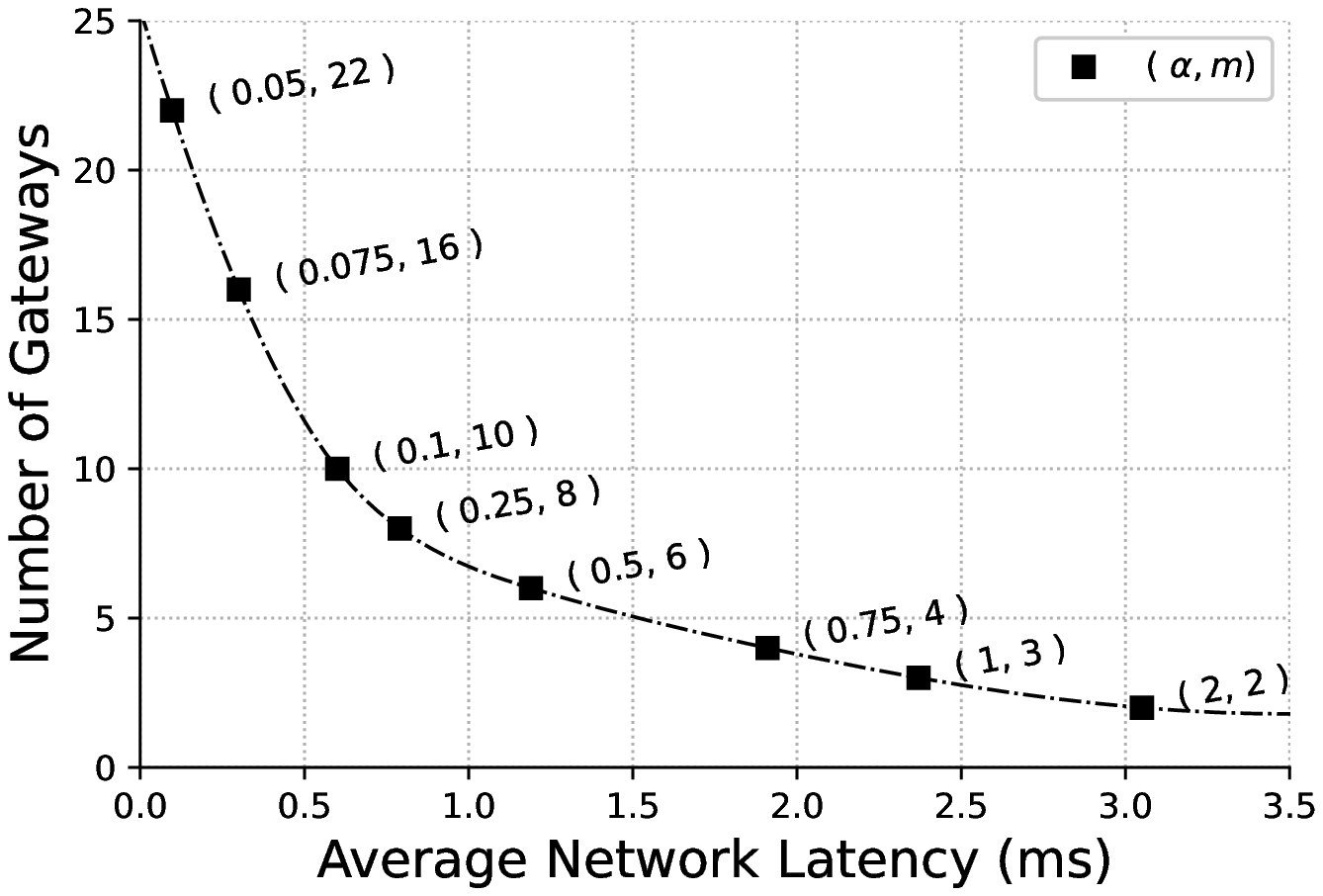}
\caption{ Exp A. Impact of $\alpha$ on the gateway deployment policy}
\label{fig:gw_tradeoff}
\end{minipage}
\end{center}
\end{figure*}

The next two experiments address the SDN controller placement problem. \textit{Exp. C} evaluates the performance of our method in jointly minimizing the synchronization cost between the SDN controllers, and the average node-to-controller latency, while \textit{Exp. D} considers the reliability-oriented SDN controller placement problem. 

Fig. \ref{fig:lat_gw_comp} compares the performance of the $(1/2)$-approximation greedy algorithm with that of the optimal MILP approach. In particular, apart from the overall objective value, the latency of the network nodes to the closest gateway is depicted in the graph, together with the number of gateways used in the resulting deployment policy. It is interesting to observe that the overall value for the objective function for the approximation method remains within $10 \%$ of the optimal MILP-based method. Moreover, the terrestrial nodes only experience at most $5\%$ increase in the tolerated latency to reach their closest gateway. 

Fig. \ref{fig:gw_tradeoff} shows how the trade-off between the two terms of the objective function can be controlled by changing the value of the parameter $\alpha$ i.e. how tuning $\alpha$ can balance the emphasis of the gateway deployment policy on minimizing the number of gateways used or minimizing the average node-to-gateway latency. For instance, lowering the value of $\alpha$ from $2$ to $1$ will almost half the amount of average node-to-gateway delay at the expense of doubling the number of deployed gateways. 

\begin{figure*}[h]
\begin{center}
\begin{minipage}[h]{0.32\textwidth}
\includegraphics[width=1\linewidth]{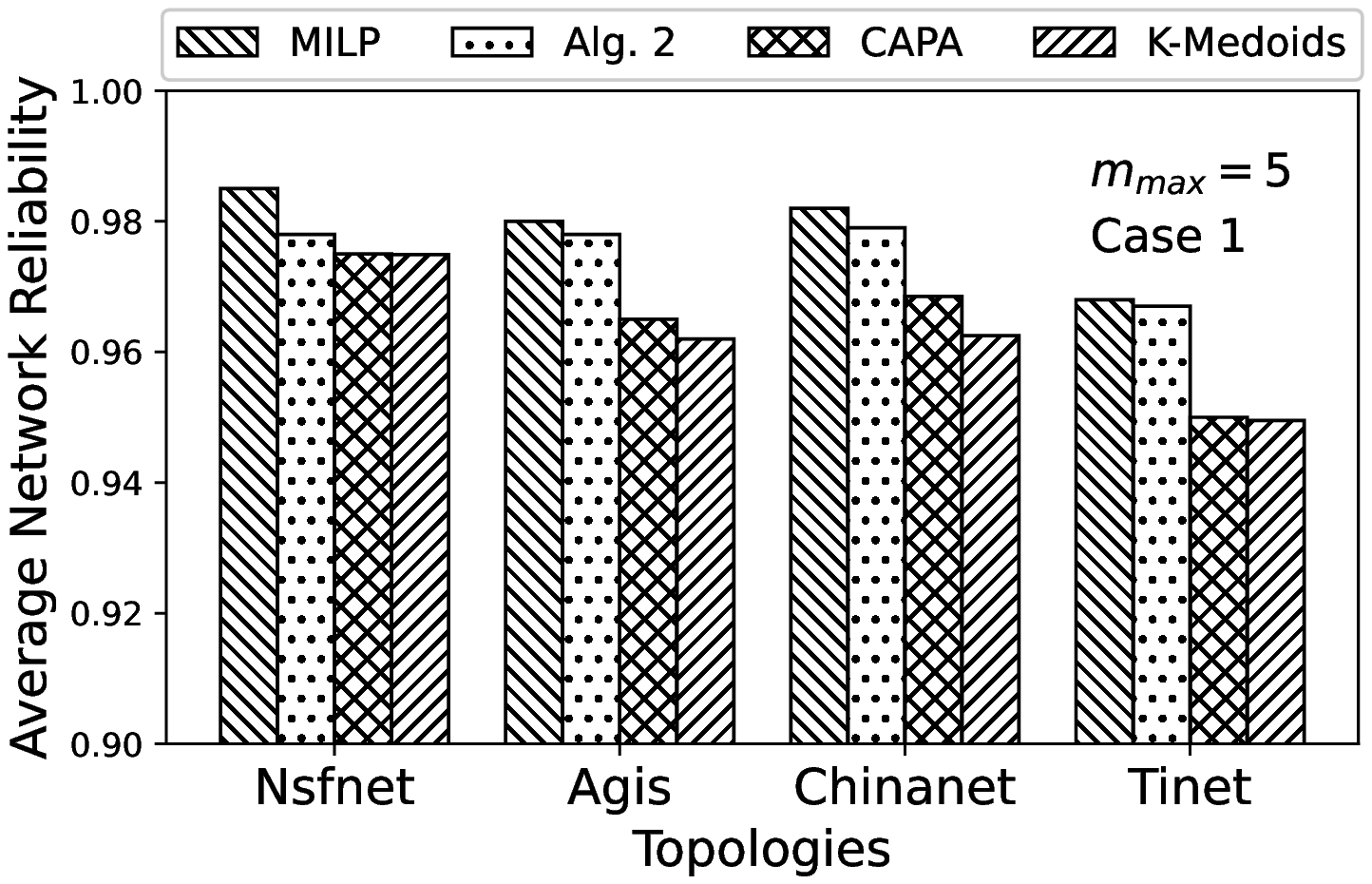}
\caption{ Exp B. Average node-to-gateway reliability}
\label{fig:rel_gw_comp}
\end{minipage}
\hspace{1mm}
\begin{minipage}[h]{0.32\textwidth}
\includegraphics[width=1\linewidth]{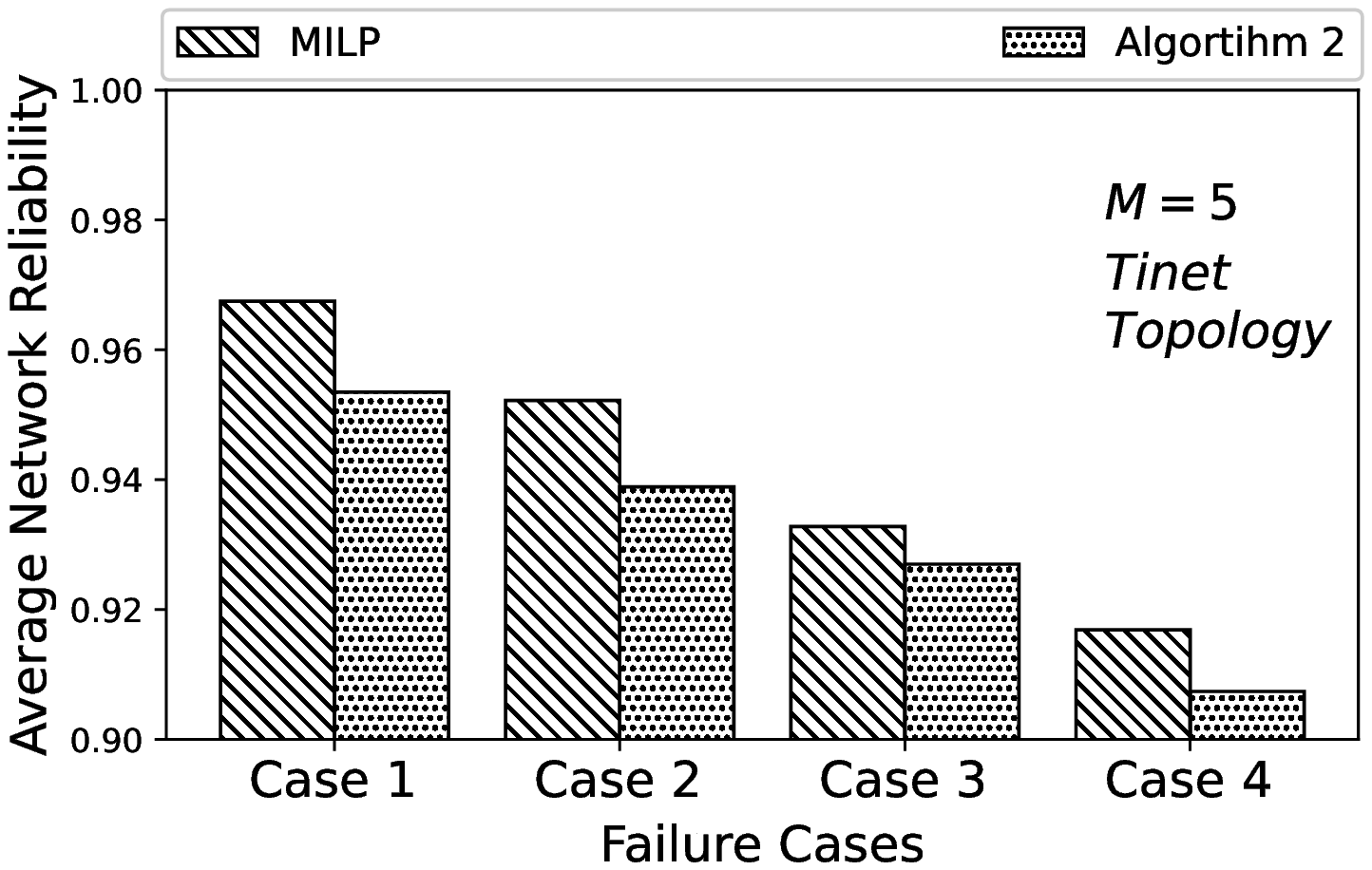}
\caption{ Exp B. Average node-to-gateway reliability under different failure prob. settings}
\label{fig:cases_gw_rel}
\end{minipage}
\hspace{1mm}
\begin{minipage}[h]{0.32\textwidth}
\includegraphics[width=1\linewidth]{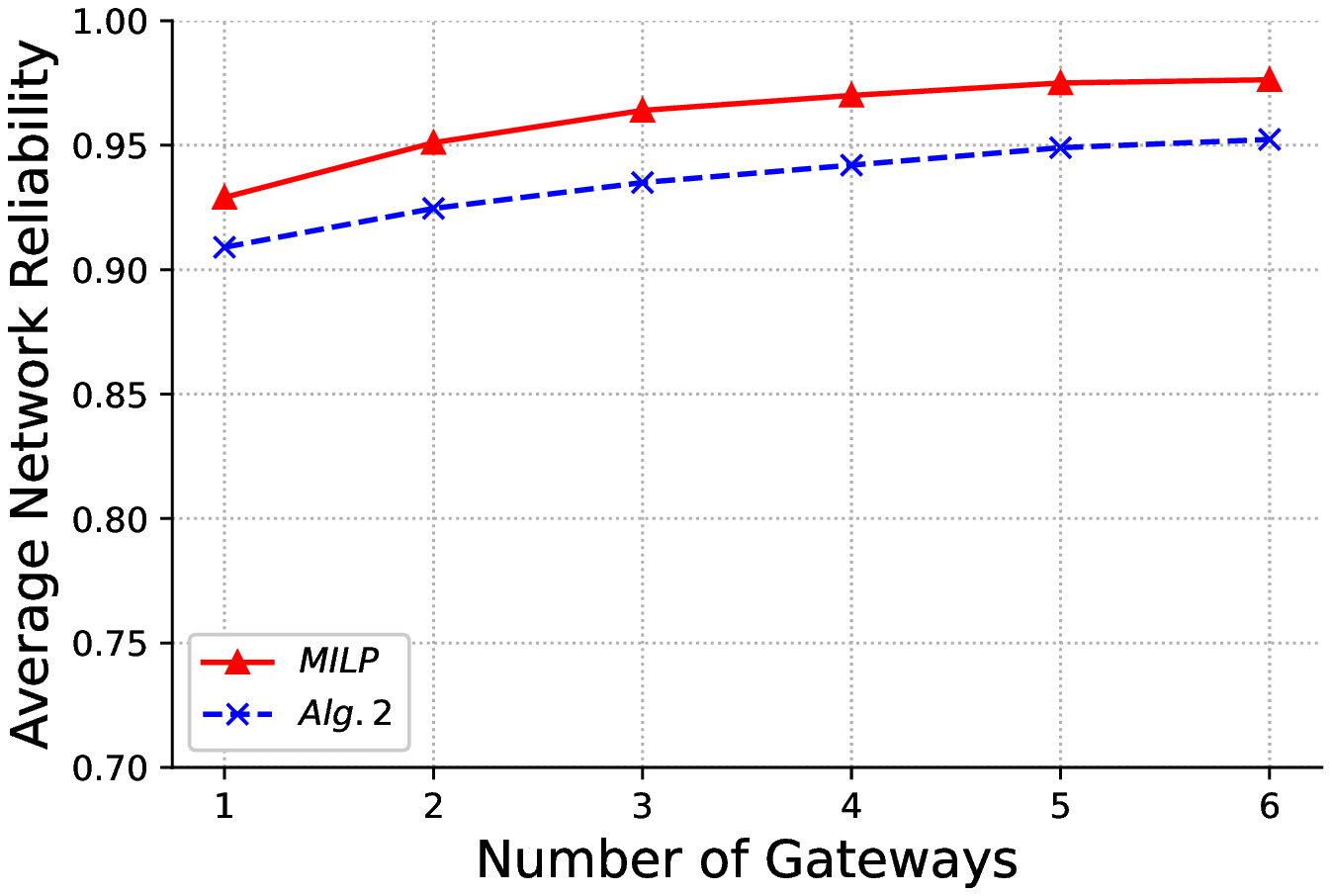}
\caption{Exp B. Impact of the number of gateways in average node-to-gateway reliability}
\label{fig:rek_change_k}
\end{minipage}
\end{center}
\end{figure*}

Fig. \ref{fig:rel_gw_comp} shows the result of running \textit{Exp. B} when a maximum of $5$ gateways is allowed to be deployed in the network. We benchmark the performance of Algorithm \ref{alg2}, in terms of average node-to-gateway reliability, against that of the approaches mentioned in \cite{onlyrel} and report the results. Similar to \cite{onlyrel}, we average our results when the failure probabilities of the network components follow \textit{Case 1} of table \ref{failure}.Our experiments show that all the methods perform reasonably well while our approach results in a performance that is slightly closer to the optimal MILP. 

With respect to different failure cases, we run \textit{Exp. B}, under $4$ different scenarios for network failure probability. In Fig. \ref{fig:rel_gw_comp}, we show the result of changing the failure settings on \textit{Tinet} topology, while the maximum number of deployed gateways cannot exceed $m_{max} = 5$.  Naturally, when the network components are more prone to failure, the average reliability decreases. It is observed that under all scenarios, the solutions generated by our approach remains close to the optimal solution. 

Fig. \ref{fig:rek_change_k}, shows the effect of changing the maximum number of gateways allowed in the gateway deployment policy. It is observed that our method follows the performance of the optimal model by at most a $3\%$ gap in terms of maximizing the average network reliability.

\begin{figure*}[h]
\begin{center}
\begin{minipage}[h]{0.32\textwidth}
\includegraphics[width=1\linewidth]{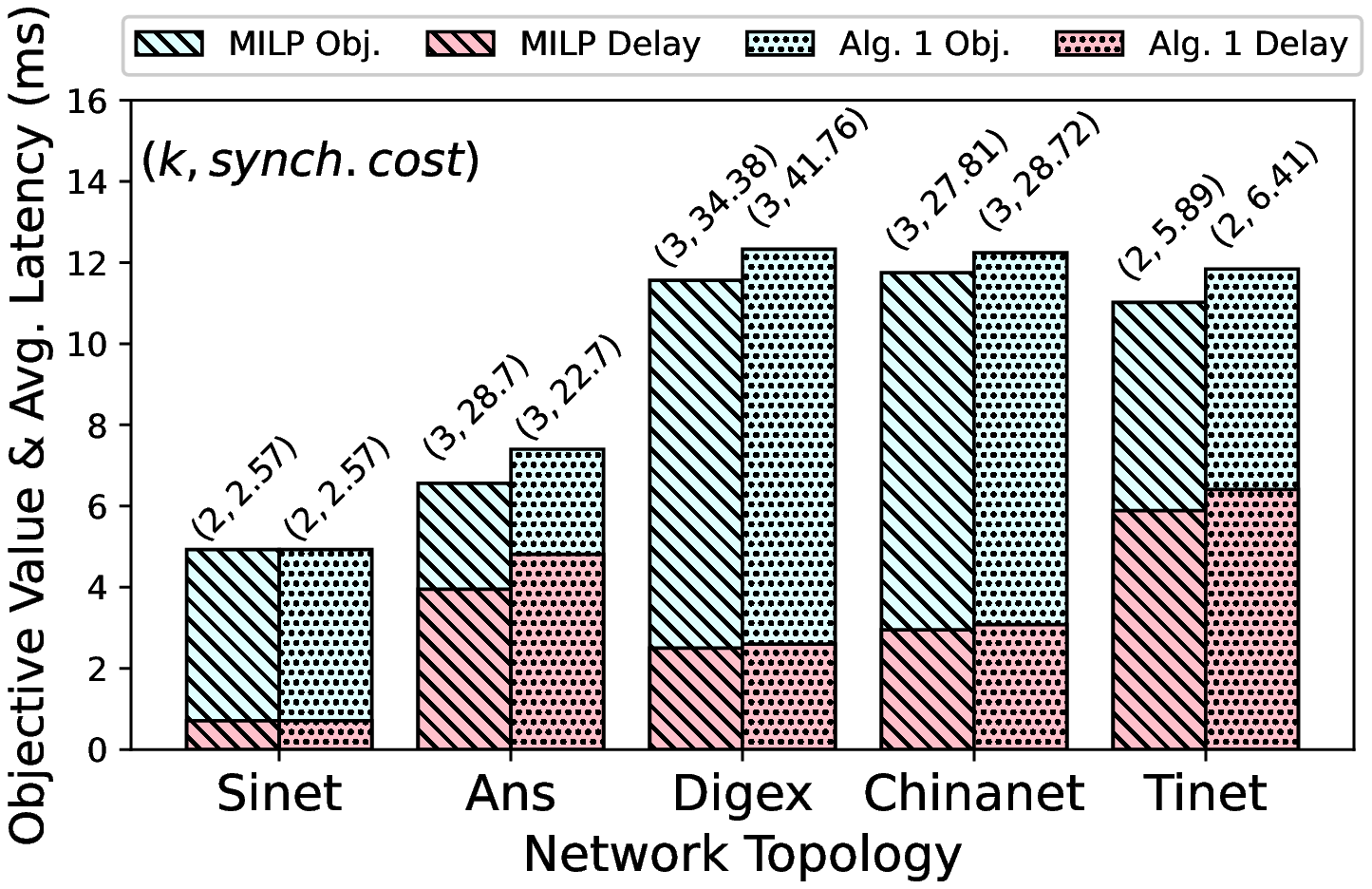}
\caption{Exp C. Objective function \& average node-to-controller latency }
\label{fig:lat_synch_con}
\end{minipage}
\hspace{1mm}
\begin{minipage}[h]{0.32\textwidth}
\includegraphics[width=1\linewidth]{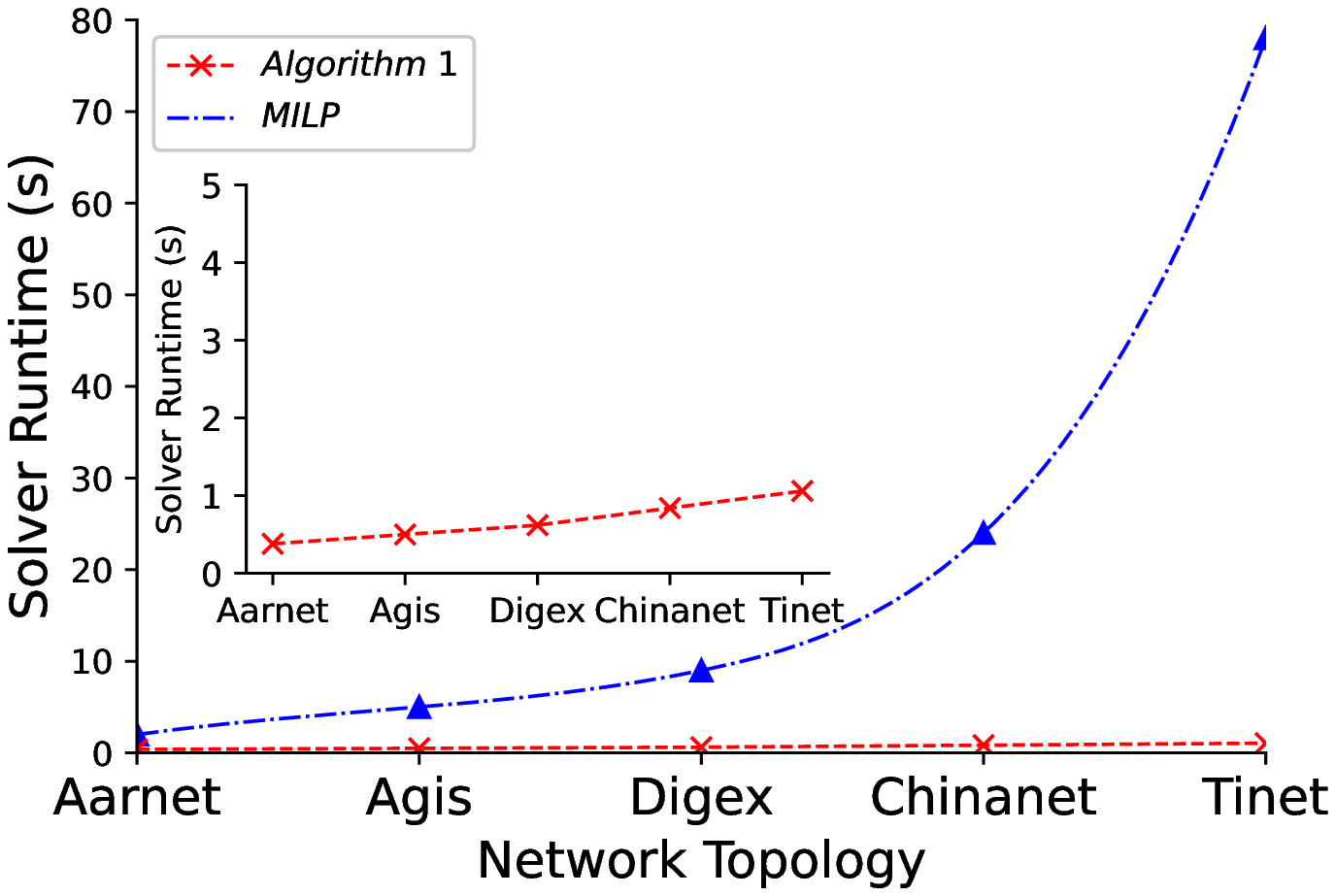}
\caption{Exp C. Comparison of Time Complexity}
\label{fig:time_complexity}
\end{minipage}
\hspace{1mm}
\begin{minipage}[h]{0.32\textwidth}
\includegraphics[width=1\linewidth]{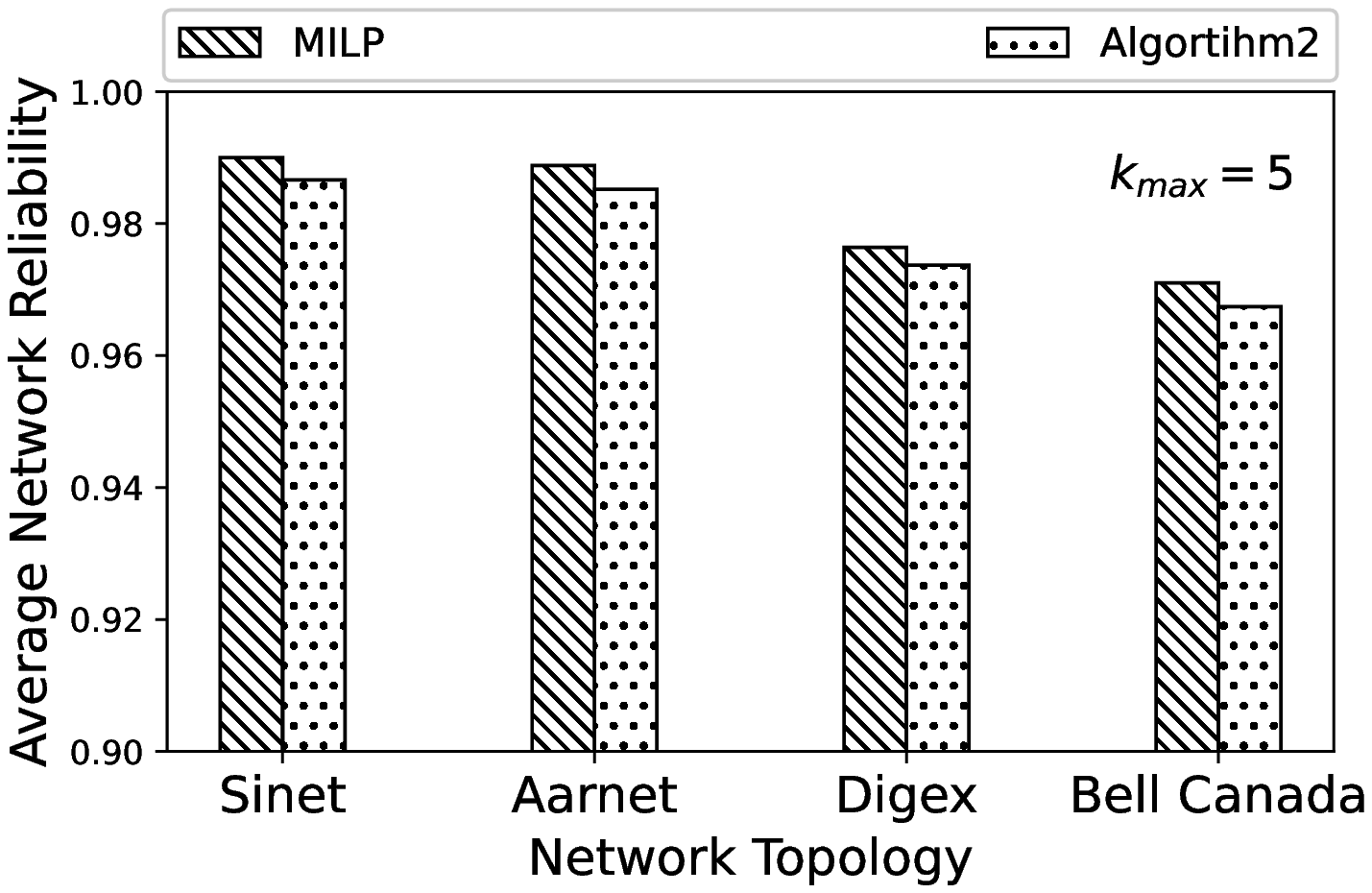}
\caption{Exp D. Average node-to-controller reliability}
\label{fig:rel_contt}
\end{minipage}
\end{center}
\end{figure*}

Fig. \ref{fig:lat_synch_con}, depicts the performance of running Exp. C, where the approximate method using Algorithm \ref{alg1} for SDN controller placement is bench-marked against the MILP-based optimal one. Our results indicate that the performance of the approximate approach remains within $10\%$ of optimality both in terms of the value of objective function and the average node-to-controller latency. The number of controllers placed and the corresponding synchronization cost is annotated on top of each topology studied. We further note that the approximate method maintains the same number of controllers as in the optimal solution. 

Fig. \ref{fig:time_complexity} illustrates the comparison between the time complexity of the MILP model (using CPLEX with branch-and-cut) and the submodular optimization algorithms. It is observed as the scale of network increases the solver runtime for solving the MILP increases exponentially, while the submodular optimization evolves linearly in terms of time which further confirms the theoretical result for its time complexity. For \textit{Tinet} topology, the MILP solver takes about $70-80$ seconds to find the optimal solution while Algorithm \ref{alg1} can run $100$ times in less than $2$ seconds. 

Finally, Fig. \ref{fig:rel_contt} shows the result of running Algorithm \ref{alg2}, for maximizing the reliability of nod-to-controller paths. The maximum number of controllers placed is enforced to not exceed $k_{max} = 5$. As our results indicate the approximate method performs reasonably well by generating near-optimal results. 

Overall, the performance evaluation experiments and analysis reveal that for all discussed variants of the satellite gateway deployment and the SDN controller placement, the approximate methods discussed in the paper can provide solutions that are very close to optimal ones in terms of accuracy, while extensively, lowering the corresponding time complexity.

%% file: conclusion.tex
\section{Conclusions}
\label{sec:conclusions}

We considered several variants of the joint satellite gateway and SDN controller placement (JGCP) in an SDN-enabled 5G-Satellite hybrid networks. We separately considered the average reliability of assignments and the latency \& overhead of the communications between SDN controllers as the objectives of the problem. We proposed a MILP model and a submodularity-based optimization framework for solving the problem, and then evaluated the performance of the proposed methods  by means of simulation. Our results confirm the effectiveness of our approach both in terms of results accuracy and time complexity. 

Within the framework of 5G-Satellite integration, the SDN controller placement problem for LEO constellations becomes more important due to the more frequent need for hand-offs between the satellite switches; and at the same time is more challenging due to the dynamically changing network topology and potentially large number of SDN controllers required in both the terrestrial and the space layer. Moreover, instead of placing the satellite gateways on the terrestrial nodes, aerial platforms can be an alternative choice leading to a cross-layer network design problem. This approach provides more flexibility to adapt with the evolving network topology, but becomes more challenging in terms of the more complicated flow routing decisions and also the partly-intermittent nature of the aerial layer communications. These two problems remain within the topics of our future research.